\documentclass[aps,prl,twocolumn,showpacs,groupedaddress,letterpaper,fleqn]{revtex4}  
\usepackage{graphicx}  
\usepackage{dcolumn}   
\usepackage{bm}        
\usepackage{amssymb}   
\usepackage{fleqn}

\newcommand{\etal}{\emph{et al.}}

\newcommand{\bone}{\ensuremath{B_1}}
\newcommand{\btwo}{\ensuremath{B_2^*}}
\newcommand{\bplus}{\ensuremath{B^+}}
\newcommand{\bst}{\ensuremath{B^{*+}}}
\newcommand{\jpsi}{\ensuremath{J/\psi}}

\begin{document}


\hspace{5.2in} \mbox{Fermilab-Pub-07/130-E}

\title{Observation and Properties of $L=1$ $\bone$ and $\btwo$ Mesons}
%
\author{V.M.~Abazov$^{35}$}
\author{B.~Abbott$^{75}$}
\author{M.~Abolins$^{65}$}
\author{B.S.~Acharya$^{28}$}
\author{M.~Adams$^{51}$}
\author{T.~Adams$^{49}$}
\author{E.~Aguilo$^{5}$}
\author{S.H.~Ahn$^{30}$}
\author{M.~Ahsan$^{59}$}
\author{G.D.~Alexeev$^{35}$}
% ...

\affiliation{\vspace{0.1 in}(The D\O\ Collaboration)\vspace{0.1 in}}
% ...
\affiliation{$^{5}$University of Alberta, Edmonton, Alberta, Canada,
                Simon Fraser University, Burnaby, British Columbia, Canada,
                York University, Toronto, Ontario, Canada, and
                McGill University, Montreal, Quebec, Canada}
% ...
\affiliation{$^{28}$Tata Institute of Fundamental Research, Mumbai, India}
% ...
\affiliation{$^{30}$Korea Detector Laboratory, Korea University, Seoul, Korea}
% ...
\affiliation{$^{35}$Joint Institute for Nuclear Research, Dubna, Russia}
% ...
\affiliation{$^{49}$Florida State University, Tallahassee, Florida 32306, USA}
% ...
\affiliation{$^{51}$University of Illinois at Chicago,
                Chicago, Illinois 60607, USA}
% ...
\affiliation{$^{59}$Kansas State University, Manhattan, Kansas 66506, USA}
% ...
\affiliation{$^{65}$Michigan State University,
                East Lansing, Michigan 48824, USA}
% ...
\affiliation{$^{75}$University of Oklahoma, Norman, Oklahoma 73019, USA}
% ...
\date{May 22, 2007}

\begin{abstract}
Excited $B$ mesons $\bone$ and $\btwo$ are observed directly for the first 
time as two separate states in fully reconstructed decays to 
$B^{+(*)} \pi^{-}$. The mass of $\bone$ is measured to be 
$5720.6 \pm 2.4 \pm 1.4$ MeV/$c^2$ and the mass difference 
$\Delta M$ between $\btwo$ and $\bone$ is $26.2 \pm 3.1 \pm 0.9$ MeV/$c^2$, giving the mass of the
$\btwo$ as $5746.8 \pm 2.4 \pm 1.7$ MeV/$c^2$.
The production rate for $\bone$ and $\btwo$ mesons is determined to be a fraction 
($13.9 \pm 1.9 \pm 3.2$)$\%$ of the production rate of the $\bplus$ meson.  
\end{abstract}


\maketitle 

To date, the detailed spectroscopy of mesons containing a $b$ quark has not been fully 
established. Only the ground $0^-$ states $\bplus$, $B^0$, $B^0_s$, $B^+_c$ and 
the excited $1^-$ state $B^*$ are considered to be established by the 
PDG \cite{pdg}. 
Quark models predict the existence of two broad ($B_{0}^*$ [$J^P = 0^+$] and
$B_{1}^*$ [$1^+$]) and two narrow ($\bone$ [$1^+$] and $\btwo$ [$2^+$]) bound
$P$ states \cite{matsuki, hqs, isgur, ebert, orsland, falk}. The broad states decay
through an $S$ wave and therefore have widths of 
a few hundred MeV/$c^2$.
Such states are difficult to distinguish, in effective mass spectra, from the combinatorial background.
The narrow states decay through a $D$ wave and therefore should have
widths of around 10 MeV/$c^2$ \cite{hqs, orsland, falk}.
Almost all observations of $\bone$ and 
$\btwo$ have been made indirectly in inclusive or semi-inclusive decays 
\cite{opal, delphi, aleph1, cdf}, which prevents their separation and a
precise measurement of their properties.
 The measurement by the ALEPH collaboration \cite{aleph2}, although partially done with 
exclusive $B$ decays, was statistically limited and model dependent. 
The masses, widths, 
and decay branching fractions of these states, in contrast, are predicted with 
good precision by various theoretical models \cite{matsuki, hqs, isgur, ebert, orsland, falk}. 
These predictions can be verified experimentally, and such a comparison 
provides important information on the quark interaction inside bound states, 
aiding further development of non-perturbative QCD. This Letter presents 
a study of narrow $L=1$ states decaying to $B^{+(*)} \pi^-$ with exclusively 
reconstructed $\bplus$ mesons using data collected by the D0 experiment
 during 2002--2006 and corresponding to an integrated luminosity of about
$1.3$ fb$^{-1}$. Throughout this Letter, charge conjugated states are implied.

The D0 detector is described in detail elsewhere~\cite{run2det}. The
detector components most important for this analysis are the central
tracking and muon systems. The D0 central tracking system consists
of a silicon microstrip tracker (SMT) and a central fiber tracker
(CFT), both located within a 2~T superconducting solenoidal magnet,
with designs optimized for tracking and vertexing at pseudorapidities
$|\eta|<3$ and $|\eta|<2.5$, respectively (where $\eta$ =
$-$ln[tan($\theta$/2)], and $\theta$ is the polar angle measured with respect to the beam line). 
The muon system is located outside the
calorimeters and has pseudorapidity coverage $|\eta|<2$.  It consists
of a layer of tracking detectors and scintillation trigger counters in
front of 1.8~T toroids, followed by two similar layers behind the
toroids~\cite{run2muon}.

The $B_1$ and $B_2^*$ mesons are studied by examining $B^{+(*)} \pi^-$ candidates.
This sample includes the following decays:
\begin{eqnarray}
\label{decay1}
\bone & \to & \bst \pi^- ;~ \bst \to \bplus \gamma ; \\
\label{decay2}
\btwo & \to & \bst \pi^- ;~ \bst \to \bplus \gamma ; \\
\label{decay3}
\btwo & \to & \bplus \pi^- .
\end{eqnarray}
The direct decay $\bone \to \bplus \pi^-$ is forbidden by 
conservation of parity and angular momentum.
The $\bplus$ meson is reconstructed in the exclusive decay 
$\bplus \to J/\psi K^+$ with $\jpsi$ decaying to $\mu^+ \mu^-$.
Each muon is required to be identified by the muon system,
have an associated track in 
the central tracking system with at least two 
measurements in the SMT, and a transverse momentum $p_T^{\mu} > 1.5$ GeV/$c$. 
At least one of the two muons is required to  
have matching track segments both inside and outside the toroidal magnet. 
The two muons must form 
a common vertex and have an invariant mass between 2.80 and 
3.35 GeV/$c^2$, to form a $\jpsi$ candidate. 
An additional charged track with $p_T > 0.5$ GeV/$c$, with total momentum 
above 0.7 GeV/$c$ and with at least two measurements in the SMT, is selected. 
This particle is assigned the kaon mass and required 
to have a common vertex, with $\chi^2 < 16$ for 3 degrees 
of freedom, with the two muons. The displacement of this vertex from the primary interaction 
point is required to exceed three standard deviations in the plane 
perpendicular to the beam direction. 
The primary vertex of the $p \bar{p}$ interaction was determined for each event
using the method described in Ref.~\cite{btag}.
The average position of the beam-collision point was included as a constraint.

From each set of three particles fulfilling these requirements, 
a $\bplus$ candidate is constructed.  
The momenta of the muons are corrected using the $\jpsi$ mass \cite{pdg} as a constraint. 
To further improve the $\bplus$ selection, a likelihood ratio method \cite{bgv} is utilized. 
This method provides a simple way of combining several discriminating variables 
into a single variable with increased power to separate signal and background.
The variables chosen for this analysis
include the smaller of the transverse momenta of the two muons, the $\chi^2$
of the $\bplus$ decay vertex, the $\bplus$ decay length divided by its
error, the significance (defined below) $S_B$ of the $\bplus$ track impact parameter,
the transverse momentum of the kaon, and the significance $S_K$ of the kaon track
impact parameter. 

For any track $i$, the significance $S_i$ is defined as 
$S_i = \sqrt{[\epsilon_T/\sigma(\epsilon_T)]^2 + 
 [\epsilon_L/\sigma(\epsilon_L)]^2}$, where $\epsilon_T$ ($\epsilon_L$)
is the projection of the track impact parameter on the plane
perpendicular to the beam direction (along the beam direction),
and $\sigma(\epsilon_T)$ [$\sigma(\epsilon_L)$] is its uncertainty.
The track of each $\bplus$ is formed assuming that 
it passes through the reconstructed vertex and is directed 
along the reconstructed $\bplus$ momentum.

\begin{figure}
    \includegraphics[width=9.3cm]
{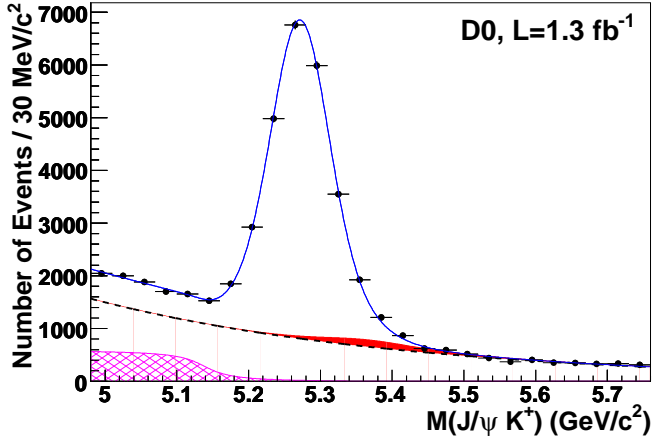}	
    \caption{Invariant mass distribution of $J/\psi K^+$ events.
        The solid line shows the sum of signal and background contributions, as described in the text.
        The contribution of $\jpsi \pi^+$ events is shown by the solid filled area and
        the $\jpsi K^{+*}$ contribution is shown by the hatched area.
        The dashed line shows the exponential function modeling the combinatorial background.}
    \label{fig1}  
\end{figure}

The resulting invariant mass distribution 
of the $\jpsi K^+$ system is shown in Fig.~\ref{fig1}. 
The curve represents the result of an unbinned likelihood fit to the sum 
of contributions from $\bplus \to \jpsi K^+$, $\bplus \to \jpsi \pi^+$, and 
$\bplus \to \jpsi K^{+*}$ decays, as well as combinatorial background. The mass distribution
of the $\jpsi K^+$ system from the $\bplus \to \jpsi K^+$ hypothesis
is parameterized by a Gaussian with the width depending 
on the momentum of the $K^+$. 
For the contribution from $\bplus \to \jpsi \pi^+$ decays, the width of
the $\jpsi \pi^+$ mass distribution is parametrized with the same
momentum-dependent width as the $\bplus \to \jpsi K^+$ decays, and then
transformed to the $\jpsi K^+$ system by assigning the kaon mass to the
charged pion.
The decay $B \to \jpsi K^{+*}$ with $K^{+*} \to K \pi$ produces a 
broad $\jpsi K^+$ mass distribution with the threshold near $M(B) - M(\pi)$. 
It is parameterized using Monte Carlo simulation (described later). 
The combinatorial background is described by an exponential function.
The $\bplus \to \jpsi K^+$ and $\bplus \to \jpsi \pi^+$ mass peaks
contain $23287 \pm 344$ (stat.) events.

\begin{figure}[htbp]
    \includegraphics[width=9.3cm]{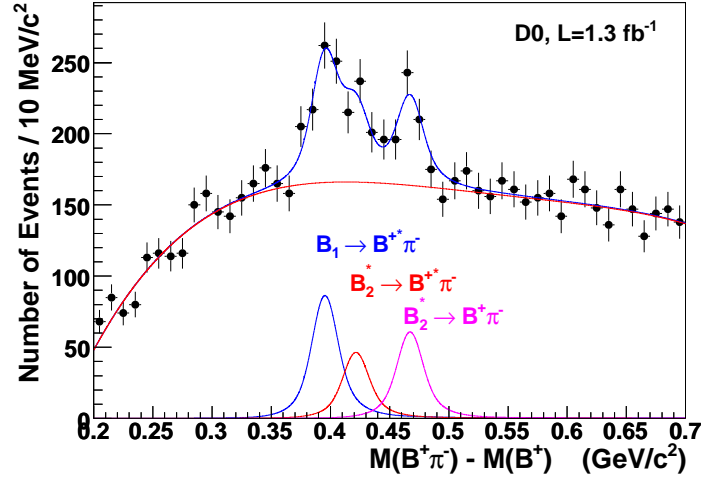}
    \caption{Invariant mass difference $\Delta M = M(B^{+} \pi^{-}) - M(B^{+})$ for 
       exclusive $B$ decays. The line shows the fit described in the text.
       The contribution of background
	and the three signal peaks are shown separately [color online].}
    \label{finalfit}  
\end{figure}

For each reconstructed $B$ meson candidate with mass
$5.19 < M(\bplus) < 5.36$ GeV/$c^2$, an additional charged
track with transverse momentum above 0.75 GeV/$c$ and  
charge opposite to that of the $B$ meson is selected.
The selection $5.19 < M(\bplus) < 5.36$ GeV/$c^2$ reduces the number of
$\bplus$ candidates to $20915 \pm 293$ (stat.).
Since the $B_J$ mesons (where $B_J$ denotes both $\bone$ and $\btwo$) decay at the production point,  
the additional track is required to originate from the primary vertex
by applying the condition on its significance $S_\pi < \sqrt{6}$.

For each combination satisfying the above criteria, the mass difference 
$\Delta M = M(B^{+} \pi^{-}) - M(B^{+})$ is computed. The resulting distribution 
of $\Delta M$ is shown in Fig.~\ref{finalfit}. The signal exhibits 
a structure that is interpreted in terms of the decays 
(\ref{decay1}--\ref{decay3}).
Since the photon from the decay $B^* \to B \gamma$ is not reconstructed,
the three decays should produce three peaks
with central positions $\Delta_1 = M(B_1) - M(B^*)$, 
corresponding to the decay $B_1 \to B^* \pi$,
$\Delta_2 = M(B_2^*) - M(B^*)$, corresponding to $B_2^* \to B^* \pi$, and
$\Delta_3 = M(B_2^*) - M(B)$, corresponding to $B_2^* \to B \pi$.
Note that in this case, $\Delta_2 = \Delta_3 - [M(\bst) - M(B)] = \Delta_3 - 45.78$ MeV/$c^2$ \cite{pdg}.
Following this expected pattern, the experimental distribution is fitted 
to the following function:
\begin{eqnarray}
F(\Delta M) & = & F_{\text{sig}}(\Delta M) + F_{\text{bckg}}(\Delta M), \nonumber \\ 
F_{\text{sig}}(\Delta M) & = & N \cdot \{ f_1 \cdot D(\Delta M, \Delta_1, \Gamma_1) 
\nonumber \\      & + & 
(1 - f_1) \cdot [ f_2 \cdot D(\Delta M, \Delta_2, \Gamma_2) \nonumber \\
& + & (1-f_2) \cdot D(\Delta M, \Delta_3, \Gamma_2)] \}.
\label{fit}
\end{eqnarray}
In these equations, $\Gamma_1$ and $\Gamma_2$ are the widths of $B_1$ and $B_2^*$, $f_1$ is the fraction of $B_1$ 
contained in the $B_J$ signal, and $f_2$ is the fraction of $B_2^* \to B^* \pi$ decays in the $\btwo$ signal. 
The parameter 
$N$ gives the total number of observed $B_J \to B^{+(*)} \pi$ decays. The background $F_{\text{bckg}}(\Delta M)$ is 
parameterized by a fourth-order polynomial.

The function $D(x, x_0, \Gamma)$ in Eq.~(\ref{fit}) 
is the convolution of a relativistic 
Breit-Wigner function with the experimental Gaussian resolution in $\Delta M$.
The width of resonances in the Breit-Wigner function takes into 
account threshold effects using the standard expression:
\begin{equation}
 \Gamma(x)   =  \Gamma_0 \frac{x_0}{x} \left(\frac{k}{k_0}\right)^{5}
 F^{(2)}(k,k_0).  \label{eqgam}
\end{equation}


The variables $k$ and $k_0$ in Eq.~(\ref{eqgam}) are the magnitudes 
of the pion three-momentum in the $B_J$ rest frame when $B_J$ has 
a four-momentum-squared equal to $x^2$ and $x_0^2$, respectively.
$\Gamma_0$ is the total decay width, and  
$F^{(2)}(k,k_0)$ is the Blatt-Weiskopf form factor for $L=2$ decay 
\cite{blwe, pdg}.

\begin{table*}[th]
\caption{\label{tab:sys1}Systematic uncertainties of the $B_J$ parameters determined from the $\Delta M$
fit. The rows show the various sources of systematic error as described in the text. The columns show the
resulting uncertainties for each of the five free signal parameters as described in Eq.~(\ref{fit}). 
$\Delta M(B_1)$ and $\Delta [M(B_2^*)-M(B_1)]$ are in MeV/$c^2$.}
\begin{ruledtabular}
\begin{tabular}{lccccr}
Source                       & $\Delta M(B_1)$ & $\Delta [M(B_2^*)-M(B_1)]$ & $\Delta R_1$ & $\Delta R_2$ & $\Delta N$ \\ \hline 
Background parameterization  & 0.15            & 0.15                       & 0.010        & 0.009        & 19         \\
Bin widths/positions         & 0.85            & 0.70                       & 0.006        & 0.026        & 12         \\
Value of $\Gamma$            & 0.75            & 0.55                       & 0.023        & 0.032        & 138        \\
$B^{+*}$ mass uncertainty    & 0.30            & 0.25                       & 0.004        & 0.004        & 6          \\
Momentum scale               & 0.50            & 0.03                       & 0.000        & 0.000        & 0          \\ 
Resolution uncertainty       & 0.20            & 0.05                       & 0.007        & 0.004        & 10         \\ 
Efficiency uncertainties     &                 &                            & 0.056        & 0.054        &            \\ \hline
Total                        & 1.30            & 0.90                       & 0.062        & 0.069        & 140        \\ 
\end{tabular}
\end{ruledtabular}
\end{table*}

The resolution in $\Delta M$ is determined from simulation.
All processes involving $B$ mesons are simulated using the {\sc EvtGen} 
generator \cite{evtgen} interfaced with {\sc pythia} \cite{pythia}, followed
by full modeling of the detector response with {\sc geant} \cite{geant} and
event reconstruction as in data. The difference between the reconstructed
and generated values of $\Delta M$ is parameterized by 
a double Gaussian function with the $\sigma$ of the narrow Gaussian set to 7.5 MeV/$c^2$, 
the $\sigma$ of the wide Gaussian set to 17.6 MeV/$c^2$, and the normalisation of the narrow Gaussian 
set to $3.8$ times that of the wide Gaussian.
Studies of the $\bplus \to \jpsi K^+$ mass peak show that simulation underestimates the mass resolution in data 
by $\approx$$10\%$. As such, the widths of the Gaussians which parameterise the $B_J$ resolution are increased
by $10\%$ to match the data, and a $100\%$ systematic uncertainty is assigned to this correction.
The widths of the observed structures are compatible
with the experimental mass resolution,
and the fit is found to be insensitive to values of 
$\Gamma_1$ and $\Gamma_2$ below the mass resolution 
with the current statistics. 
Therefore, both these widths are fixed at 10 MeV/$c^2$ in the fit, 
as suggested by theoretical models \cite{hqs, falk}. 
They are varied together over a wide range to estimate the associated systematic 
uncertainty.

With these assumptions, the following parameters of $B_1$ and $B_2^*$ are obtained:
\begin{eqnarray}
M(B_1) - M(\bplus) & = & 441.5 \pm 2.4 \pm 1.3~\mbox{MeV/}c^2, \nonumber \\
M(B_2^*) - M(B_1)  & = & 26.2 \pm 3.1 \pm 0.9~\mbox{MeV/}c^2. 
\label{masses}
\end{eqnarray}
where the first uncertainty is statistical, and the second is systematic.
The correlation coefficient of these mass measurements is $-0.659$. 
With these relations, and using the mass of the $\bplus$ \cite{pdg}, the absolute
masses of the $B_1$ and $B_2^*$ are:
\begin{eqnarray}
M(B_1)       & = & 5720.6 \pm 2.4 \pm 1.4~\mbox{MeV/}c^2, \nonumber \\
M(B_2^*)     & = & 5746.8 \pm 2.4 \pm 1.7~\mbox{MeV/}c^2. 
\label{absmasses}
\end{eqnarray}

The number of $B_J$ decays is found to be $N = 662 \pm 91$. 
The $\chi^2/$d.o.f. of the fit is $33/40$.  
Without the $B_J$ signal 
contribution, the $\chi^2/$d.o.f. of the fit increases to $97/45$, which implies 
that this structure is observed with a statistical significance 
of more than $7 \sigma$. Fitting with only one peak increases 
the $\chi^2/$d.o.f. to $54/42$, which corresponds to more than a $4 \sigma$ significance 
that more than one resonance is observed. 
With the $\btwo \to \bst \pi$ decay removed from the fit, the $\chi^2/$d.o.f. of the fit
increases to $41/41$. Although with the current statistics we can not distinguish
between the two- and three- peaks hypotheses, theory suggests that
$\btwo$ decays with almost equal branching ratios into $B \pi$ and $\bst \pi$ \cite{hqs, falk},
and our fit indeed indicates a preference for this expected pattern.

The number of $B_J$ mesons and values $f_1$ and $f_2$ obtained from the fit 
are used to measure the production and decay ratios of $\bone$ and $\btwo$:
\begin{eqnarray}
R_1 & = & \frac{Br(B_1 \to \bst \pi)}{Br(B_J \to B^{(*)} \pi)}  =   f_1 \cdot \frac{\varepsilon_0}{\varepsilon_1} \nonumber \\
    & = & 0.477 \pm 0.069 \pm 0.062, \nonumber \\
R_2 & = & \frac{Br(\btwo \to B^* \pi)}{Br(\btwo \to B^{(*)} \pi)}  =   f_2 \cdot \frac{\varepsilon_3}{\varepsilon_2} \nonumber \\
    & = & 0.475 \pm 0.095 \pm 0.069, \nonumber \\
R_J & = & \frac{Br(b \to B_J^0 \to B^{(*)} \pi)}{Br(b \to B^+)}  =  \frac{3 \cdot N(B_J)}{2 \cdot N(\bplus) \cdot \varepsilon_0} \nonumber \\
    & = & 0.139 \pm 0.019 \pm 0.032. 
\label{rates}
\end{eqnarray}

Here $\varepsilon_1$, $\varepsilon_2$ and $\varepsilon_3$ are the efficiencies to select an additional pion
from the $B_J$ decay for decay modes $\bone \to \bst \pi^-$, $\btwo \to \bst \pi^-$ and
$\btwo \to \bplus \pi^-$, respectively.
They are determined from a simulation separately for each decay mode (\ref{decay1}--\ref{decay3}).
The overall efficiency for detecting a pion from any $B_J \to B^{+(*)} \pi^-$ decay is
$\varepsilon_0 = 0.342 \pm 0.008 \pm 0.028$. 
The value for $R_J$ takes into account the decay $B_J \to B^0 \pi^0$ assuming isospin conservation.
%

For the $B_J$ mass fit, the influences of different sources of systematic uncertainty 
are estimated by examining the changes in the fit parameters under a number of variations. 
Different background parameterizations are used in the fit to the $\Delta M$ distribution. 
In addition, the effect of binning is tested by varying the bin width
and position. 
The parameters describing the background are allowed to vary in the fit and their uncertainties 
are included in our results. To check the effect of fixing 
$\Gamma_1$ and $\Gamma_2$ at 10 MeV/$c^2$, a range of widths from 0 to 20 MeV/$c^2$ is 
used.
The effect of the uncertainty on the mass difference 
$M(B^{+*}) - M(B^+)$ \cite{pdg} is also taken into account.
Different parameterizations of the detector mass resolution are tested, and in addition 
the fit is made without the $10\%$ mass resolution correction. 
The uncertainty in the absolute momentum scale, which results in a small shift
of all measured masses, is also taken into account.
The summary of all systematic uncertainties in the $B_J$ mass fit is given in Table~\ref{tab:sys1}.  

%
The measurement of the relative production rate $R_J$ uses 
the pion detection efficiencies predicted in simulation, 
as well as the numbers of $B_J$ and $\bplus$ events.
To estimate the systematic uncertainty on the number of $\bplus$ events, 
different parameterizations of the signal and background are used for the fit. 
The resulting uncertainty is $\pm 200$ $B^+$ events.
The systematic uncertainty on the number of $B_J$ events is $\pm 140$ (see Table~\ref{tab:sys1}). 
The uncertainty of the impact parameter resolution in the simulation
is estimated to be $\approx$$10\%$ \cite{burdin2}.
It can influence the measurement of the selection efficiency of the pion
from the $B_J$ decay, and its contribution to the systematic uncertainty
of $R_J$ is found to be 0.0056.
The track reconstruction efficiency for particles with low transverse
momentum is measured in Ref.~\cite{brat} and
good agreement between data and simulation is found. 
This comparison is valid within the uncertainties of branching fractions
of different $B$ semileptonic decays, which is about 7\%. This uncertainty
results in a 0.0096 variation of $R_J$.
 An additional systematic uncertainty of 0.0008 associated with the
difference in the momentum distributions of selected particles
in data and in simulation is taken into account.
Combining all these effects in quadrature, the total systematic uncertainty
in the relative production rate $R_J$ is found to be 0.032, 
of which the dominant contribution comes from the uncertainty 
on the number of $B_J$ events.

Different consistency checks of the observed signal are performed. 
The stability of the fit 
under different selections of $\pi$ meson from $B_J \to B^{+(*)} \pi$
decay and the width of the $\bplus$ mass window is verified.
Events with positively and negatively charged pions are analyzed 
separately, and consistent results are obtained.
A complementary sample of events containing a pion not compatible 
with the primary vertex is selected, and no significant $B_J$ signal 
is observed. Events with wrong charge combinations 
($B^{+}\pi^{+}$ and $B^{-}\pi^{-}$) also show a signal consistent 
with zero. 
In addition, the fit is repeated without the Blatt-Weiskopf form-factor and 
no visible change in results is observed.

In conclusion, the $B_1$ and $B_2^*$ mesons are observed for the first time 
as two separate states. Their measured masses  
are given by Eq.~(\ref{masses}).
The $B_J$ production rate,
the branching fraction of $B_2^*$ to the excited state $B^*$,
and the fraction of the $B_1$ meson in the $B_J$  production rate
are also measured as given in Eq.~(\ref{rates}). These results
will help to develop models describing bound states with heavy quarks.

%
We thank the staffs at Fermilab and collaborating institutions, 
and acknowledge support from the 
DOE and NSF (USA);
CEA and CNRS/IN2P3 (France);
FASI, Rosatom and RFBR (Russia);
CAPES, CNPq, FAPERJ, FAPESP and FUNDUNESP (Brazil);
DAE and DST (India);
Colciencias (Colombia);
CONACyT (Mexico);
KRF and KOSEF (Korea);
CONICET and UBACyT (Argentina);
FOM (The Netherlands);
PPARC (United Kingdom);
MSMT (Czech Republic);
CRC Program, CFI, NSERC and WestGrid Project (Canada);
BMBF and DFG (Germany);
SFI (Ireland);
The Swedish Research Council (Sweden);
Research Corporation;
Alexander von Humboldt Foundation;
and the Marie Curie Program.
%


\begin{thebibliography}{99}

%
\bibitem[*]{alton}
Visitor from Augustana College, Sioux Falls, SD, USA.
\bibitem[\P]{burdin}
Visitor from The University of Liverpool, Liverpool, UK.
\bibitem[\S]{podesta-lerma}
Visitor from ICN-UNAM, Mexico City, Mexico.
\bibitem[\ddag]{voutilainen}
Visitor from Helsinki Institute of Physics, Helsinki, Finland.
\bibitem[\#]{wenger}
Visitor from Universit{\"a}t Z{\"u}rich, Z{\"u}rich, Switzerland.
%
\vskip 0.25cm

\bibitem{pdg} W.-M. Yao \etal\ (Particle Data Group),     J. Phys. G {\bf 33}, 1 (2006).

\bibitem{matsuki}
  T.~Matsuki and T.~Morii,                               Phys.\ Rev.\ D {\bf 56}, 5646 (1997);

  T.~Matsuki, T.~Morii and K.~Sudoh, arXiv:hep-ph/0605019.

\bibitem{hqs} 
  M.~Di Pierro and E.~Eichten, Phys.\ Rev.\ D {\bf 64}, 114004 (2001);
  E.J.~Eichten, C.T.~Hill, C.~Quigg, Phys.\ Rev.\ Lett. {\bf 71}, 4116 (1993).

\bibitem{isgur} N.~Isgur,                                Phys.\ Rev.\ D {\bf 57}, 4041 (1998).

\bibitem{ebert} D.~Ebert, V.O.~Galkin, R.N.~Faustov,     Phys.\ Rev.\ D {\bf 57},
5663 (1998), Erratum                                     Phys.\ Rev.\ D {\bf 59}, 019902 (1999).

\bibitem{orsland} A.H.~Orsland, H.~Hogaasen,             Eur.\ Phys.\ J. {\bf C9}, 503 (1999).

\bibitem{falk} A.~Falk, T.~Mehen,			 Phys.\ Rev.\ D {\bf 53}, 231 (1996).

\bibitem{opal} OPAL Collaboration, R.~Akers \etal,       Z.\ Phys.\ C {\bf 66}, 19 (1995).

\bibitem{delphi} DELPHI Collaboration, P.~Abreu \etal,   Phys.\ Lett.\ B {\bf 345}, 598 (1995).

\bibitem{aleph1} ALEPH Collaboration, D.~Buskulic \etal, Z.\ Phys.\ C {\bf 69}, 393 (1996).

\bibitem{cdf} CDF Collaboration, T.~Affolder \etal,      Phys.\ Rev.\ D {\bf 64}, 072002 (2001).

\bibitem{aleph2} ALEPH Collaboration, R.~Barate \etal,   Phys.\ Lett.\ B {\bf 425}, 215 (1998).

\bibitem{run2det} D0 Collaboration, V.M.~Abazov \etal,   Nucl.\ Instrum.\ Methods A {\bf 565}, 463 (2006).

\bibitem{run2muon} D0 Collaboration, V.M.~Abazov \etal,  Nucl.\ Instrum.\ Methods A {\bf 552}, 372 (2005).

\bibitem{btag} DELPHI Collaboration, J.~Abdallah \etal,  Eur.\ Phys.\ J {\bf C32}, 185 (2004).

\bibitem{bgv} G.~Borisov,                                Nucl.\ Instrum.\ Methods A {\bf 417}, 384 (1998).

\bibitem{blwe} J.~Blatt and V.~Weisskopf, Theoretical Nuclear Physics (John Wiley \& Sons, New York, 1952), p. 361.

\bibitem{evtgen} D.J.~Lange,                             Nucl.\ Instrum.\ Methods A {\bf 462}, 152 (2001).

\bibitem{pythia} T.~Sj\"ostrand \etal,                   Comp.\ Phys.\ Commun. {\bf 135}, 238 (2001).

\bibitem{geant} R.~Brun and F.~Carminati,                CERN Program Library Long Writeup {\bf W5013} (1993).

\bibitem{brat} D0 Collaboration, V.M.~Abazov \etal,      Phys.\ Rev.\ Lett. {\bf 94}, 182001 (2005).

\bibitem{burdin2} D0 Collaboration, V.M.~Abazov \etal,    Phys.\ Rev.\ Lett. {\bf 97}, 021802 (2006).

\end{thebibliography}
\end{document}